\documentclass[aps,showpacs,onecolumn,notitlepage,floatfix,amsmath,amssymb,nofootinbib]{revtex4-1}

\usepackage{amssymb}
\usepackage{amsmath}
\usepackage{graphicx,float}
\usepackage{subfig}
\usepackage{color}
\usepackage{latexsym}
\usepackage{bm,latexsym}
\usepackage{mathrsfs}
\usepackage{pbox}
\usepackage{bigints} 
\usepackage[normalem]{ulem}
\usepackage[usenames,dvipsnames]{xcolor}
\definecolor{orange}{cmyk}{0,0.5,1,0}

\selectfont{}
\begin{document}

\title{Dyonic Simpson Visser black-bounce spacetime surrounded by an exotic Einstein cluster
in general relativity}

\author{Pedro Ca\~nate$^{1}$}
\email{pcannate@gmail.com}

\author{Joseph Sultana$^{2}$}
\email{joseph.sultana@um.edu.mt}

\author{Demosthenes Kazanas$^{3}$}
\email{dkazanas51@gmail.com, demosthenes.kazanas@nasa.gov}

\affiliation{$^{1}$Departamento de Ciencias Naturales, Facultad de Ciencias Exactas y Naturales, Universidad Surcolombiana, Avenida Pastrana Borrero - Carrera 1, A.A. 385, Neiva, Huila, Colombia.\\
$^{2}$Department of Mathematics, Faculty of Science, University of Malta, Msida, Malta.\\
$^{3}$Astrophysics Science Division, Code 663, NASA Goddard Space Flight Center, Greenbelt, MD 20771, USA}

\begin{abstract}
In this work, we present the Simpson-Visser black-bounce (SV-BB) spacetime 
with both electric and magnetic charges (the dyonic SV-BB) as an exact solution of general relativity coupled to Maxwell electrodynamics and an anisotropic fluid with vanishing radial pressure.
We assume the anisotropic fluid represents an Einstein cluster, which consists of a 
dense system of particles moving in circular orbits that generate only tangential pressure and zero radial pressure. We show that the electromagnetic SV-BB metric, which provides a regular black hole spacetime, can be obtained as an exact solution to the Einstein field equations sourced by a combination of an Einstein cluster (which is responsible for providing the exotic matter required to keep a black-bounce throat open) and an electromagnetic field in the framework of Maxwell electrodynamics.
\end{abstract}

\pacs{04.20.Jb, 04.50.Kd, 04.50.-h, 04.40.Nr}


\maketitle

\section{Introduction}

Regular black holes are an interesting class of singularity-free spacetimes that plays an important role in the study of quantum-corrected black holes (see \cite{quantum_BH} for details).

Motivated by the Bardeen model \cite{Bardeen1968}
several regular black holes, in the sense that none of them have curvature singularities,\footnote{I.e., all invariants
constructed from the Riemann curvature tensor $\boldsymbol{\mathcal{R}}=R_{\alpha\beta\mu\nu}\boldsymbol{dx^{\alpha}\!\otimes\! dx^{\beta}\!\otimes\!dx^{\mu}\!\otimes\!dx^{\nu}}$, e.g.  $R \equiv R^{\alpha}{}_{\alpha}$, $R_{\alpha\beta}R^{\alpha\beta}$, $R_{\alpha\beta\mu\nu}R^{\alpha\beta\mu\nu}$ and similar scalars formed by polynomial expressions in covariant derivatives of the curvature tensor,
are well-defined everywhere.} have been constructed in the framework of the Einstein's theory of general relativity coupled to nonlinear electrodynamics (GR-NLED) \cite{Ayon_Garcia,Ayon2000}. 
Moreover, in Ref. \cite{Ayon_Garcia1999}, a new model of regular black hole solution (which is not of the Bardeen type) was constructed within the framework of GR-NLED. 
However, these models are either purely electric \cite{Ayon_Garcia,Ayon_Garcia1999}, with electromagnetic invariant $\mathcal{F}\sim -\mathcal{E}^{2}$
(where $\mathcal{E}$ denotes the electric field vector), or purely magnetic \cite{Ayon2000} with $\mathcal{F}\sim \mathcal{B}^{2}$ (where $\mathcal{B}$ denotes the magnetic field vector). 
Indeed, the purely electric regular black hole solutions \cite{Ayon_Garcia} can be considered 
as particular cases of the Pellicer-Torrence solution with an electric charge \cite{Pellicer69}. \\
Although these solutions (purely electric or purely magnetic) obey the weak energy condition (WEC), they can evade the singularity theorems \cite{Penrose,Penrose_GC,Wald2,Senovilla}
because they are not globally hyperbolic. It is important to note that because the regular black hole metrics derived in \cite{Ayon_Garcia,Ayon_Garcia1999} are formulated with solely electric charges, this necessitates distinct NLED Lagrangians in different regions of spacetime; this shortcoming was introduced in Refs. \cite{Bronnikov2000,Bronnikov2001}.\\ 
Simpson and Visser (SV) \cite{Simpson_Visser} proposed a new class of singularity-free black hole, known as a black bounce (BB), which, in contrast to Bardeen black holes,
smoothly interpolates between the Schwarzschild black hole and the Ellis-Bronnikov traversable wormhole \cite{morris88,morris88-2}, and is described by the static, spherically symmetric and asymptotically flat (SSS-AF) metric given by,
\begin{equation}\label{B_bounce}
\boldsymbol{ds^{2}} = - \left( 1 - \frac{2m}{\sqrt{r^{2} + a^{2}}} \right)\boldsymbol{dt^{2}} +  \left( 1 - \frac{2m}{ \sqrt{r^{2} + a^{2} } } \right)^{-1} \boldsymbol{dr^{2}}
+ \left( r^{2} + a^{2} \right)\boldsymbol{d\Omega^{2}},
\end{equation}
where $m$ and $a$ are real parameters; $t$ and $r$ are the temporal and radial coordinates, respectively, while $\left( r^{2} + a^{2} \right)\boldsymbol{d\Omega^{2}}$, with $\boldsymbol{d\Omega^{2}}= \boldsymbol{d\theta^{2}} + \sin^{2}\theta \boldsymbol{d\varphi^{2}}$, is the line element of a two-dimensional sphere of radius $\sqrt{r^{2} + a^{2}}$.
The ranges of these coordinates are $t\in(-\infty,\infty)$, $r\in(-\infty,\infty)$, $\theta\in[0,\pi]$ and $\phi\in[0,2\pi)$.

For this metric the associated curvature invariants are given by,
\begin{eqnarray}
&&R = \frac{ 2 a^{2} (3m - \sqrt{r^{2} + a^{2}} ) }{ \left(r^{2} + a^{2}\right)^{^{\frac{5}{2}}} }, \quad \quad\quad\quad\quad\quad\quad %
R_{\alpha\beta}R^{\alpha\beta} = \frac{a^{4}\left[   4\left(\sqrt{r^{2} + a^{2}} - \frac{3}{2}m \right)^{2} + (3m)^{2} \right] }{ \left(r^{2} + a^{2}\right)^{^{5}} }, \label{CI1_R}\\
&&R_{\alpha\beta\mu\nu}R^{\alpha\beta\mu\nu} = \frac{4 \left[8a^{2}m(r^{2} - a^{2})\sqrt{r^{2} + a^{2}} + 3a^{4}(r^{2} + a^{2}) + 3m^{2}( 3a^{4}-4a^{2}r^{2} + 4r^{4}) \right] }{\left(r^{2} + a^{2}\right)^{^{5}}}. \label{CI2_R}
\end{eqnarray}
which, for $a\neq0$, are well defined for all real values of $r$. Furthermore, in \cite{Simpson_Visser}, the metric given in Eq.~(\ref{B_bounce})
is presented as a minimal modification of the Schwarzschild spacetime. By adjusting the parameter $a$, this metric represents different spacetimes. For $a=0\neq m$, it reduces to the ordinary Schwarzschild spacetime, whereas for
$a\neq0=m$, it reduces to the ordinary Ellis-Bronnikov traversable wormhole spacetime.
For $a\neq0\neq m$, the nature of the spacetime depends on the relation between
$m$ and $a$. 
{\it Simpson-Visser regular black hole configuration:} if $2m\geqslant|a|$, the spacetime represents a regular black hole with an event
horizon at $r=\sqrt{4m^{2}-a^{2}}$, since the curvature invariants
\eqref{CI1_R}--\eqref{CI2_R} are well defined for all real values of the radial coordinate
$r$, and with a black-bounce throat at $r=0$ connecting the
asymptotically flat regions $r\in(-\infty,0)$ and $r\in(0,\infty)$. 
{\it Simpson-Visser traversable wormhole configuration:} if $2m<|a|$, no event horizon exists and the spacetime becomes a traversable
wormhole in the Morris-Thorne sense, with a wormhole throat at $r=0$.  

In this paper, we demonstrate a novel way of realizing the SV-BB spacetime as an exact solution of the Einstein field equations, in contrast to the works \cite{Bronnikov2022,Canate2022,Rodrigues2023} that presented the SV-BB as an exact solution of the Einstein field equations supported by a purely magnetic nonlinear electromagnetic field and a self-interacting phantom scalar field. 
We show that, within the framework of general relativity (GR), this spacetime can instead be sourced by a linear electromagnetic field ({\it i.e.}, one that satisfies the Maxwell equations) and an anisotropic fluid with vanishing radial pressure representing an Einstein cluster. 
The concept of an Einstein cluster (EC) was introduced in \cite{Einstein1938} to represent 
a self gravitating system formed by a large number of non-interacting particles (cluster) 
moving on circular geodesics in all possible angular directions and orbital radii around the center of symmetry of the cluster. Therefore, an EC constitutes a static and spherically symmetric system of particles, where the angular momentum of each particle in the cluster generates an outward centrifugal force that counteracts the gravitational attraction experienced by the particle.
As a result of this, an EC has an anisotropic fluid description with zero radial pressure, non-zero tangential pressure and positive energy density.\\
Einstein clusters have been extensively examined in the literature, with a special emphasis on the astrophysical context in which they have been presented as potential models of galactic dark matter halos (for a recent review, see \cite{Maeda2025,Acharyya2024} and references therein).
\\
Einstein cluster models also have been used to describe the interior Schwarzschild solution \cite{Florides,Boehmer}; as models of inhomogeneous spacetimes \cite{Szybka2020}; and compact star mimickers \cite{Singh2019}.
\\
Furthermore, the spherical gravitational collapse of a compact object under the approximation that the radial pressure is identically zero has been discussed in \cite{Singh1997,Mahajan2007}.

This paper is structured as follows. 
The previously proposed interpretation of the Simpson-Visser black-bounce spacetime as a purely magnetic solution of Einstein-nonlinear electrodynamics gravity coupled to a self-interacting scalar field is reviewed in the next section.
In Sec.~\ref{BBs-ED-FL}, within the framework of general relativity, we introduce a new source for the Simpson-Visser black-bounce spacetime.
In Sec.~\ref{Uniqueness}, we show that the Simpson-Visser black-bounce solution for the Einstein-Maxwell theory coupled to an exotic Einstein cluster is unique.
The conclusions are presented in the final section. In this paper we use units where $G = k_{_{B}} = c =\hbar = 1$, and the metric signature $(-+++)$ is used throughout.

\section{General relativity coupled to nonlinear electrodynamics and additional matter fields }\label{BBs-ST-NLED}

General relativity coupled to nonlinear electrodynamics (GR-NLED) and additional matter fields is defined by the action
\begin{equation}\label{action_STNLED}
\mathcal{S}[g_{ab},A_{a},\psi_{a} ] = \frac{1}{16\pi}\int
\left[R + 4\mathcal{L}(\mathcal{F}) \right]\sqrt{-g} d^{4}x
+ \mathcal{S}_{\!_{\mathrm{matt}}}[g_{ab},\psi_{a}],
\end{equation}
where $R$ is the scalar curvature,
$\mathcal{L}=\mathcal{L}(\mathcal{F})$ is a function of the electromagnetic invariant $\mathcal{F}\equiv \frac{1}{4}F_{\alpha\beta}F^{\alpha\beta}$, where $F_{\alpha\beta}=2\partial_{[\alpha}A_{\beta]}$ are
the components of the electromagnetic field tensor  $\boldsymbol{F}=\frac{1}{2}F_{\alpha\beta} \boldsymbol{dx^{\alpha}} \wedge \boldsymbol{dx^{\beta}}$ and $A_{\alpha}$ are the components of the electromagnetic potential. 
The first term is the Einstein-Hilbert action minimally coupled to NLED, while the second term $\mathcal{S}_{\!_{\mathrm{matt}}}[g_{ab},\psi_{a}]=\frac{1}{16\pi}\int \mathcal{L}_{\!_{\mathrm{matt}}}(\psi_{a}) \sqrt{-g} d^{4}x$
is the action for the additional matter, where $\mathcal{L}_{\!_{\mathrm{matt}}}(\psi_{a})$ represents the Lagrangian density and $\psi_{a}$ the additional matter fields.
For the case of purely GR-NLED, {\it i.e.,} the action \eqref{action_STNLED} with $\mathcal{S}_{\!_{\mathrm{matt}}}[g_{ab},\psi_{a}]=0$, in Ref. \cite{Ayon2000}, the Bardeen black hole solution  \cite{Bardeen1968} ~--the first regular black hole model in general relativity--~ was reinterpreted as a purely magnetic solution to Einstein
equations coupled to a nonlinear electrodynamics. The metric that describes the Bardeen black hole, in the framework of GR-NLED, is as follow
\begin{equation}\label{RBH_Bard}
ds^{2} =  - \left( 1 - \frac{2mr^{2}}{\sqrt{\left(r^{2} + q^{2}_{_{\mathcal{B}}}\right)^{3}}} \right) dt^{2} + \left( 1 - \frac{2mr^{2}}{\sqrt{\left(r^{2} + q^{2}_{_{\mathcal{B}}}\right)^{3}}} \right)^{-1}dr^{2}  + r^{2}d\Omega^{2},
\end{equation}
where $q_{_{\mathcal{B}}}$ is the magnetic charge of a self-gravitating magnetic
field described by a nonlinear electrodynamics model, which depends only on the electromagnetic invariant $\mathcal{F}$, given by
\begin{equation}\label{LF_RBH_Bard}
 \mathcal{L}(\mathcal{F}) = \frac{3}{2sq^{2}_{_{\mathcal{B}}}}\!\left( \frac{ \sqrt{2q^{2}_{_{\mathcal{B}}}\mathcal{F}} }{1 + \sqrt{2q^{2}_{_{\mathcal{B}}}\mathcal{F}} }\right)^{\!\!^{\frac{5}{2}}},
\end{equation}
where $s$ stands for $s=|q_{_{\mathcal{B}}}|/(2m)$, and for the NLED Lagrangian to be real-valued, $\mathcal{F}$ must be a positive quantity
which, in this case, is given by $\mathcal{F} = \frac{q^{2}_{_{\mathcal{B}}}}{2r^{4}}$.
In other words, the spacetime metric \eqref{LF_RBH_Bard} is a purely magnetic solution of the GR-NLED field equations \eqref{Eqt_NDSF}-\eqref{phi2_NDSF} in Appendix, with $\mathcal{L}(\mathcal{F})$ given by \eqref{LF_RBH_Bard} and $\phi=0=\mathscr{U}$.
\\
Einstein-nonlinear electrodynamics gravity coupled to a self-interacting scalar field (GR-NLED-SF) provides an interesting case of \eqref{action_STNLED} with  non-vanishing $\mathcal{S}_{\!_{\mathrm{matt}}}[g_{ab},\psi_{a}]$,
representing the action of a self-interacting scalar field $\phi$, minimally coupled to gravity.
Thus, the matter Lagrangian density is given by
\begin{equation}\label{phi_matt}
\mathcal{L}_{\!_{\mathrm{matt}}}(\psi_{a})=
\mathcal{L}\!_{_{_{S \! F }}}(\phi) = - \frac{1}{2}g^{\mu\nu}\nabla_{\mu}\phi\nabla_{\nu}\phi  - 2 \mathscr{U}(\phi),  \end{equation}
where the potential $\mathscr{U}(\phi)$ represents the self-interaction of $\phi$.
The GR-NLED-SF field equations are
\begin{equation}\label{ES_NLED_Eqs}
G_{\alpha}{}^{\beta} = 8\pi (T_{\alpha}{}^{\beta})\!_{_{_{N \! L \! E \! D}}} \!+ \! 8\pi (T_{\alpha}{}^{\beta})\!_{_{_{S \! F}}},
\end{equation}
where $G_{\alpha}{}^{\beta} = R_{\alpha}{}^{\beta} - \frac{R}{2} \delta_{\alpha}{}^{\beta}$ denotes the components of the Einstein tensor, $(T_{\alpha}{}^{\beta})\!_{_{_{N \! L \! E \! D}}}$  are the components of the NLED energy-momentum tensor,
\begin{equation}\label{NLED_EM}
8\pi(T_{\alpha}{}^{\beta})\!_{_{_{N \! L \! E \! D}}} = 2\mathcal{L}_{_{\mathcal{F}}}F_{\alpha\mu}F^{\beta\mu} - 2\mathcal{L}\hskip.06cm\delta_{\alpha}{}^{\beta}, 
\end{equation}
where $\mathcal{L}_{_{\mathcal{F}}}\equiv \frac{d\mathcal{L}}{d\mathcal{F}}$ and
with electromagnetic field tensor $F_{\mu\nu}$ fulfilling the source-free modified Maxwell’s equations
\begin{equation}
\nabla_{\mu}(\mathcal{L}_{_{\mathcal{F}}}F^{\mu\nu}) = 0 = d\boldsymbol{F},
\end{equation}
whereas $(T_{\alpha}{}^{\beta})\!_{_{_{S \! F}}}$ are the components of the energy-momentum tensor of self-interacting scalar field,
\begin{equation}\label{E_SF}
8 \pi(T_{\alpha}{}^{\beta})\!_{_{_{S \! F}}} = -\frac{1}{4}(\nabla_{\mu}\phi \hspace{0.04cm} \nabla^{\mu}\phi)\delta_{\alpha}{}^{\beta} + \frac{1}{2}\nabla_{\alpha}\phi \hspace{0.04cm} \nabla^{\beta}\phi - \mathscr{U} \hspace{0.03cm} \delta_{\alpha}{}^{\beta},
\end{equation}
with $\phi$ satisfying the equation of motion
\begin{equation}\label{FEq_sf}
\nabla_{\mu}\!\nabla^{\mu}\phi = 2\hspace{0.03cm}\dot{\mathscr{U}},
\end{equation}
with $\dot{\mathscr{U}} = \frac{d\mathscr{U}}{d\phi}$.\\ 
In Appendix, we state the GR-NLED-SF field equations for a generic static, spherically symmetric and purely magnetic spacetime.

\subsection{Simpson-Visser black-bounce as a purely magnetic exact solution of the GR-NLED-SF field equations}

In Refs.~\cite{Bronnikov2022,Canate2022,Rodrigues2023}, it was shown that the following power-law Maxwell NLED Lagrangian and scalar potential, given respectively by,
\begin{equation}\label{U_L_SV}
\mathcal{L}(\mathcal{F}) = s_{_{0}}|\mathcal{F}|^{^{\!\frac{5}{4}}}
\quad\quad\textup{ and }\quad\quad
\mathscr{U}(\phi) = \mathscr{U}_{_{0}}\hspace{0.04cm}\cosh^{5}\left(\frac{\phi}{2}\right),
\end{equation}
where $s_{_{0}}$  and $\mathscr{U}_{_{0}}$ are real parameters of the theory, defines a NLED-SF model for which the pure magnetically charged SV-BB given by,
\begin{equation}\label{Qm_SV_BB}
\boldsymbol{ds^{2}} = - \left( 1 - \frac{2m}{\sqrt{ r^{2} + q^{2}_{_{\mathcal{B}}} } } \right)\boldsymbol{dt^{2}} +  \left( 1 - \frac{2m}{ \sqrt{r^{2} + q^{2}_{_{\mathcal{B}}} } } \right)^{-1} \boldsymbol{dr^{2}}
+ \left( r^{2} + q^{2}_{_{\mathcal{B}}} \right)\boldsymbol{d\Omega^{2}},
\end{equation}
with $q_{_{\mathcal{B}}}$ and $m$ being real parameters, representing the magnetic charge and mass, respectively, of the SV-BB, is an exact solution of the GR-NLED-SF field equations \eqref{Eqt_NDSF}-\eqref{phi2_NDSF} in Appendix. In this case the scalar field is given by
\begin{equation}\label{sf_SV}
\phi(r) = 2i\tan^{^{\!\!-1}}\!\!\left(\frac{r}{q_{_{\mathcal{B}}}}\right),
\end{equation}
and the electromagnetic field tensor is given by
\begin{equation}\label{Fab_SSS_pmf}
F_{\alpha\beta} =  \Big( \delta^{\theta}_{\alpha}\delta^{\varphi}_{\beta} - \delta^{\varphi}_{\alpha}\delta^{\theta}_{\beta}\Big) F_{\theta\varphi} \quad\quad \textup{ with } \quad\quad F_{\theta\varphi} = q_{_{\mathcal{B}}} \sin\theta,
\end{equation}
with parameters $(s_{_{0}}, \mathscr{U}_{_{0}}, q_{_{\mathcal{B}}}, m)$ that satisfy
$s_{_{0}}=\frac{6}{5}\frac{(2)^{^{\frac{1}{4}}}m}{\sqrt{|q_{_{\mathcal{B}}}|} }$ and $\mathscr{U}_{_{0}} = \frac{4}{5}\frac{m}{|q_{_{\mathcal{B}}}|^{3}}$.
\\
{\bf Avoiding the Penrose singularity theorem:} For the purely magnetic SV-BB \eqref{Qm_SV_BB},
with the power-law Maxwell NLED \eqref{U_L_SV} 
\begin{equation}
\mathcal{F} = \frac{q^{2}_{_{\mathcal{B}}}}{2\left( r^{2} + q^{2}_{_{\mathcal{B}}} \right)^{2}} >0, \quad\quad\quad \mathcal{L} = s_{_{0}}\mathcal{F}^{^{\frac{5}{4}}} > 0  \quad\quad\textup{ and }\quad\quad \mathcal{L}_{\mathcal{F}} = \frac{5s_{_{0}}}{4}\mathcal{F}^{^{\frac{1}{4}}} > 0  \quad\quad\textup{ for all }\quad\quad
r\in\mathbb{R}.
\end{equation}
According to \eqref{WEC_NLED}, it follows that the NLED field source of the SV-BB satisfies the null and weak energy conditions, and hence the local energy density of the nonlinear electrodynamic field $\rho^{^{(loc)}}_{\!_{_{NLED}}}= (T_{\mu\nu})\!_{_{_{N\!L\!E\!D}}}k^{\mu}k^{\nu}$
({\it i.e.,} the NLED energy density measured by an observer with a timelike 4-velocity
 $k^{\mu}$) must be non-negative everywhere for all timelike $k^{\mu}$.
An important requirement for an NLED theory is that it
reduce to linear electrodynamics in the weak-field limit
theory [also known as the Maxwell weak-field limit; $\mathcal{L}(\mathcal{F}) \sim \mathcal{F}$ and $\mathcal{L}_{_{\mathcal{F}}}(\mathcal{F})\sim 1$ as $\mathcal{F}\sim0$].
We can note that the NLED model $\mathcal{L}(\mathcal{F}) = s_{_{0}}|\mathcal{F}|^{^{\frac{5}{4}}}$
does not satisfy this requirement.
\\
On the other hand, for the scalar field \eqref{sf_SV}, we can observe that $\phi(r)$
is imaginary throughout the entire SV-BB spacetime; hence, this is a phantom scalar field.
Furthermore, for the total matter source of the SV-BB [{\it i.e.,} the energy-momentum tensor that represents the combination of the scalar field and the NLED $(T_{\alpha}{}^{\beta})\!_{_{_{e\!f\!f\!}}}=(T_{\alpha}{}^{\beta})\!_{_{_{S \! F }}} + (T_{\alpha}{}^{\beta})\!_{_{_{N \! L \! E \! D}}}$],
using Eqs.~\eqref{sf_SV}, \eqref{WEC_SF}, \eqref{WEC_SF_Tr}, \eqref{WEC_effect} and \eqref{WEC_effectTr}, for the Simpson-Visser spacetime \eqref{Qm_SV_BB}, the following is obtained:
\begin{eqnarray}
 && 8\pi\rho\!_{_{_{e\!f\!f}}} + 8\pi P\!_{_{_{e\!f\!f}}} = 8\pi\rho\!_{_{_{S \! F }}} + 8\pi P\!_{_{_{S \! F }}} = -\frac{2q^{2}_{_{\mathcal{B}}}}{\left( r^{2} + q^{2}_{_{\mathcal{B}}} \right)^{2}}\!\left(1-\frac{2m}{\sqrt{ r^{2} + q^{2}_{_{\mathcal{B}}} }}\right) \quad\textup{for}\quad 2m < \sqrt{ r^{2} + q^{2}_{_{\mathcal{B}}} }, \label{E_SF_SVBB} \\
 && 8\pi\rho\!_{_{_{e\!f\!f}}} + 8\pi P\!_{_{_{e\!f\!f}}} = 8\pi\rho\!_{_{_{S \! F }}} + 8\pi P\!_{_{_{S \! F }}} = \frac{2q^{2}_{_{\mathcal{B}}}}{\left( r^{2} + q^{2}_{_{\mathcal{B}}} \right)^{2}}\!\left(1-\frac{2m}{\sqrt{ r^{2} + q^{2}_{_{\mathcal{B}}} }}\right)
\quad\textup{for}\quad 2m > \sqrt{ r^{2} + q^{2}_{_{\mathcal{B}}} }, \label{E_SF_SVBB_Tr}
\end{eqnarray}
Thus, we can deduce that for the SV regular black hole configuration $2m \geqslant |q_{_{\mathcal{B}}}|$, 
the quantity $(\rho\!_{_{_{e\!f\!f}}} + P\!_{_{_{e\!f\!f}}})$ is negative for all $|r| \in \mathbb{R}^{+} - \left\{ |r_{_{h}}|=\sqrt{4m^{2} - q^{2}_{_{\mathcal{B}}}} \right\}$.  
This corresponds to the entire SV regular black hole spacetime, with the exception of horizons $r=r_{_{h}}$ where $(\rho\!_{_{_{e\!f\!f}}} + P\!_{_{_{e\!f\!f}}})\big|_{_{r=r_{_{h}}}}= 0$ due to $g_{_{tt}}(r_{_{h}}) = 0$.\\
Whereas, for the SV traversable wormhole configuration $2m<|q_{_{\mathcal{B}}}|$,  
yields $2m < \sqrt{ r^{2} + q^{2}_{_{\mathcal{B}}} }$ for all $r\in\mathbb{R}$. Hence, according to Eq. \eqref{E_SF_SVBB}, the quantity $(\rho\!_{_{_{e\!f\!f}}} + P\!_{_{_{e\!f\!f}}})$ is negative-definite 
across the SV traversable wormhole spacetime.\\
Thus, when interpreted \eqref{Qm_SV_BB} as a purely magnetic exact solution of GR-NLED-SF, the solution does not satisfy the WEC, with the self-interacting scalar field being solely responsible for the WEC violation, and therefore the gravitational collapse leading to the Simpson-Visser black-bounce spacetime evades the singularity theorem of Penrose \cite{Penrose}. See Appendix for details. 

In the context of GR-NLED-SF, the construction of dyonic black-bounces of the Simpson-Visser type was recently discussed in \cite{Junior2025}. 
They used the inverse method, which consists of starting from a target black-bounce metric, employing the consistency of GR-NLED-SF field equations to calculate the NLED Lagrangian and the scalar field potential required to generate that metric (see Eqs. (19), (22), and (23) of \cite{Junior2025}). 
However, although this approach allows one to find a NLED-SF system for which a specific dyonic black-bounce metric satisfies the GR-NLED-SF field equations, it turns out that none of the generated NLED models can be considered physically reasonable because they do not reduce to Maxwell electrodynamics in the weak-field limit
(see equations (42), (58), (61), (63), (71), (84), and (87) of \cite{Junior2025}). Furthermore, 
these NLED models necessitate such fine tuning that their NLED Lagrangians are determined by the electric and magnetic charges of their corresponding target dyonic black-bounce solutions. 

\section{Einstein field equations for static and spherically symmetric spacetime
with a Maxwell electromagnetic field and an anisotropic fluid as sources}
\label{BBs-ED-FL}

General relativity coupled to Maxwell electrodynamics and an anisotropic fluid (GR-Maxwell-FL) is defined by the action \eqref{action_STNLED}, for the case of Maxwell electrodynamics $\mathcal{L}(\mathcal{F}) = \mathcal{F}$, and  with $\mathcal{S}_{\!_{\mathrm{matt}}}[g_{ab},\psi_{a}]$ representing an anisotropic fluid (FL).
The GR-Maxwell-FL field equations arising from action \eqref{action_STNLED} are,
\begin{eqnarray}
&&G_{\alpha}{}^{\beta} = 8\pi (T_{\alpha}{}^{\beta})\!_{_{_{\mathrm{ED}}}} \! + \! 8\pi (T_{\alpha}{}^{\beta})\!_{_{_{\mathrm{FL}}}},  \label{ES_NLED_Eqs} \\
&&\nabla_{\mu} F^{\mu\nu} = 0 = d\boldsymbol{F},  \label{ES_NLED_Eqs2} \\
&&\nabla_{\mu} (T_{\nu}{}^{\mu})\!_{_{_{\mathrm{FL}}}} = 0, \label{ES_NLED_Eqs3}
\end{eqnarray}
where $(T_{\alpha}{}^{\beta})\!_{_{_{\mathrm{ED}}}}$  are the components of the linear electrodynamics (ED) energy-momentum tensor,
\begin{equation}\label{ED_EM}
8\pi(T_{\alpha}{}^{\beta})\!_{_{_{\mathrm{ED}}}} = 2F_{\alpha\mu}F^{\beta\mu} - 2\mathcal{F}\hskip.06cm\delta_{\alpha}{}^{\beta}. %
\end{equation}
Whereas,  $(T_{\alpha}{}^{\beta})\!_{_{_{\mathrm{FL}}}}$ are the components of the 
energy-momentum tensor for the anisotropic fluid
\begin{equation}\label{T_FL}
(T_{\alpha}{}^{\beta})\!_{_{_{\mathrm{FL}}}} = (\rho\!_{_{_{\mathrm{FL}}}} + \tilde{P}\!_{_{_{\mathrm{FL}}}}) u_{\alpha}{}u^{\beta} + \tilde{P}\!_{_{_{\mathrm{FL}}}} g_{\alpha}{}^{\beta} + (P\!_{_{_{\mathrm{FL}}}} - \tilde{P}\!_{_{_{\mathrm{FL}}}})v_{\alpha}{}v^{\beta},
\end{equation}
where $\rho\!_{_{_{\mathrm{FL}}}}$, $P\!_{_{_{\mathrm{FL}}}}$, and $\tilde{P}\!_{_{_{\mathrm{FL}}}}$ are the comoving energy density, radial pressure, and tangential pressure of the anisotropic fluid, respectively. Here, $u_{\alpha}$ is the four-velocity of the fluid, and $v_{\alpha}$ is a spacelike vector in the radial direction, which satisfy the conditions: $u_{\alpha}u^{\alpha}=-1$, $v_{\alpha}v^{\alpha}=1$ and $u_{\alpha}v^{\alpha}=0$.\\
We now express the field equations \eqref{ES_NLED_Eqs}$-$\eqref{ES_NLED_Eqs3} for the case of a static spherically symmetric (SSS)  configuration.
We assume that the metric takes the static and spherically symmetric form
\begin{equation}\label{SSSmet}
\boldsymbol{ds^{2}} =  -  A(r)\boldsymbol{dt^{2}} + \frac{\boldsymbol{dr^{2}} }{A(r)}  + \chi^{2}(r) \boldsymbol{d\Omega^{2}},
\end{equation}
written in standard spherical coordinates $(t, r, \theta, \varphi)$,
where $A=A(r)$ and $\chi=\chi(r)$ are functions that depend only on the radial coordinate $r$.\\
For the metric ansatz \eqref{SSSmet} the non-null components of the Einstein tensor are given by,
\begin{equation}\label{GabSSS}    
G_{t}{}^{t}\!=\!\frac{ \left( 2\chi\chi'' + \chi'^{2} \right)A + \chi\chi' A' - 1  }{ \chi^{2} }, \hspace{0.35cm}  G_{r}{}^{r}\!=\!\frac{ \chi'^{2}A + \chi\chi' A' - 1  }{ \chi^{2} } \hspace{0.35cm}\textup{and}\hspace{0.35cm} G_{\theta}{}^{\theta}\!=\! G_{\varphi}{}^{\varphi}\!=\!\frac{ 2\left(\chi''A + \chi'A' \right) + \chi A'' }{ 2\chi }.
\end{equation}
For the electromagnetic field tensor $F_{\alpha\beta}$, because the spacetime metric \eqref{SSSmet} is static and spherically symmetric, the only nonvanishing terms are the electric $F_{tr}$ and magnetic $F_{\theta\phi}$ components, so that $F_{\alpha\beta}$ can be written as
\begin{equation}\label{Fab_SSS}
F_{\alpha\beta} = \Big( \delta^{t}_{\alpha}\delta^{r}_{\beta} - \delta^{r}_{\alpha}\delta^{t}_{\beta}\Big) F_{tr} + \Big( \delta^{\theta}_{\alpha}\delta^{\varphi}_{\beta} - \delta^{\varphi}_{\alpha}\delta^{\theta}_{\beta}\Big) F_{\theta\varphi}.
\end{equation}
Hence, the nonvanishing components of $(T_{\alpha}{}^{\beta})\!_{_{_{\mathrm{ED}}}}$
obtained from the metric given in Eq.\!~\eqref{SSSmet}, with the electromagnetic field tensor in Eq.\!~\eqref{Fab_SSS}, are given by,
\begin{eqnarray}
&& 8\pi (T_{t}{}^{t})\!_{_{_{\mathrm{ED}}}} = 8\pi (T_{r}{}^{r})\!_{_{_{\mathrm{ED}}}} = 2(F_{tr}F^{tr} - \mathcal{F}), \label{NLEDtt}\\
&&8\pi (T_{\theta}{}^{\theta})\!_{_{_{\mathrm{ED}}}} = 8\pi (T_{\varphi}{}^{\varphi})\!_{_{_{\mathrm{ED}}}} = 2(F_{\theta\varphi}F^{\theta\varphi} - \mathcal{F}). \label{NLEDtheph}
\end{eqnarray}  
Using Eqs.\!~\eqref{ES_NLED_Eqs2}, written as  $\partial_{\alpha}(\sqrt{-g} F^{\alpha\beta}) = 0 = d(F_{\alpha\beta} \boldsymbol{dx^{\alpha}} \wedge \boldsymbol{dx^{\beta}})$, we obtain
\begin{equation}\label{F_comps}
F^{rt} = \frac{ q_{_{\mathcal{E}}} }{\chi^{2}\sqrt{-g_{tt}g_{rr}}~ }, \quad\quad\quad\quad\quad
F^{\theta\varphi} = \frac{q_{_{\mathcal{B}}}}{\chi^{4} \sin\theta },
\end{equation}
where $q_{_{\mathcal{E}}}$ and $q_{_{\mathcal{B}}}$ are the electric and magnetic charges of the spacetime, respectively.
Thus, the nonvanishing components of $F_{\alpha\beta}$  and the invariant $\mathcal{F}$ for the generic
static and spherically symmetric spacetime geometry Eq.\!~\eqref{SSSmet}, are given by
\begin{equation}\label{F_invariant}
F_{rt} = -\frac{ \sqrt{-g_{tt}g_{rr} } ~ q_{_{\mathcal{E}}}  }{\chi^{2}},\quad\quad\quad  F_{\theta\varphi} = q_{_{\mathcal{B}}} \sin\theta, \quad\quad\quad \mathcal{F} = \frac{q^{2}_{_{\mathcal{B}}}  -  q^{2}_{_{\mathcal{E}}}}{2\chi^{4}}.
\end{equation}
Hence, Eq.~\eqref{F_invariant} gives the most general solution of  
Maxwell equations \eqref{ES_NLED_Eqs2}
for static and spherically symmetric spacetimes.
Substituting Eqs. \eqref{F_comps} and \eqref{F_invariant} into the Eqs. \eqref{NLEDtt} and \eqref{NLEDtheph}, we obtain  
\begin{eqnarray}
&& 8\pi(T_{t}{}^{t})\!_{_{_{\mathrm{ED}}}} = 8\pi(T_{r}{}^{r})\!_{_{_{\mathrm{ED}}}} 
= - \frac{q^{2}}{\chi^{4}}, \label{EDttSSS}\\
&& 8\pi(T_{\theta}{}^{\theta})\!_{_{_{\mathrm{ED}}}} =  8\pi(T_{\varphi}{}^{\varphi})\!_{_{_{\mathrm{ED}}}}
= \frac{q^{2}}{\chi^{4} } . \label{EDthephSSS}
\end{eqnarray}
where $q^{2} = q^{2}_{_{\mathcal{B}}}  +  q^{2}_{_{\mathcal{E}}}$. 
The dependence of the electromagnetic stress-energy tensor on the charges only through this combination follows from the invariance of source-free Maxwell electrodynamics under electric–magnetic duality rotations. Consequently, the gravitational field equations and the resulting metric depend on the effective charge ($q$), rather than on the separate electric and magnetic charges. (see \cite{Gibbons95,Bronnikov26,Bandos} for review). 
\\
On the other hand, for the generic SSS spacetime geometry \eqref{SSSmet}, 
with respect to a comoving observer with 4-velocity
$\boldsymbol{u} = - A^{^{-\frac{1}{2}}} \boldsymbol{\partial_{t}}$,
and $\boldsymbol{v} = - A^{^{\frac{1}{2}}} \boldsymbol{\partial_{r}}$,
the component of the energy-momentum tensor for the fluid
takes the form
\begin{equation}\label{T_DF}
(T_{\alpha}{}^{\beta})\!_{_{_{\mathrm{FL}}}} = - \rho ~\delta_{\alpha}{}^{t}\delta_{t}{}^{\beta} + P ~\delta_{\alpha}{}^{r}\delta_{r}{}^{\beta} + \tilde{P} ~(\delta_{\alpha}{}^{\theta}\delta_{\theta}{}^{\beta} + \delta_{\alpha}{}^{\varphi}\delta_{\varphi}{}^{\beta} ),
\end{equation}
Here, $\rho, P$, and $\tilde{P}$ are functions of the radial coordinate $r$, representing the energy density, radial pressure, and tangential pressure of the anisotropic fluid, respectively. In the $R$-regions, where $A(r)>0$, the metric has signature $(-+++)$, and $t$ and $r$ are the temporal and spatial coordinates\footnote{According to the notation of Ref. \cite{Bronnikov2022}}, respectively. In these regions, $\rho$ and $P$ represent the comoving energy density and radial pressure of the fluid. In the $T$-regions, where $A(r)<0$, the metric has signature $(+-++)$, and the roles of $t$ and $r$ are interchanged: $r$ becomes the temporal coordinate and $t$ the spatial coordinate. Consequently, in the $T$-regions, the comoving energy density and radial pressure are given by $-P$ and $-\rho$, respectively.
Thus, for the metric ansatz \eqref{SSSmet},
inserting the components \eqref{NLEDtt}, \eqref{NLEDtheph} and  \eqref{T_DF}
in the field equations \eqref{ES_NLED_Eqs} written as
$\mathscr{C}_{\alpha}{}^{\beta} \equiv G_{\alpha}{}^{\beta} - 8\pi [ (T_{\alpha}{}^{\beta})\!_{_{_{\mathrm{ED}}}} + (T_{\alpha}{}^{\beta})\!_{_{_{\mathrm{FL}}}}
] = 0$, we obtain 
\begin{eqnarray}
&&\!\mathscr{C}_{t}{}^{t}\!=\!0\hspace{1.4cm}\!\Rightarrow\!\hspace{0.4cm}
\frac{ \left( 2\chi\chi'' + \chi'^{2} \right)A + \chi\chi' A' - 1  }{ \chi^{2} } 
+ \frac{q^{2}}{\chi^{4}}
+ 8\pi \rho
= 0, \label{Eqt}\\
&&\nonumber\\
&& \!\mathscr{C}_{r}{}^{r}\!=\!0\hspace{1.4cm}\!\Rightarrow\!\hspace{0.4cm} \frac{ \chi'^{2}A + \chi\chi' A' - 1  }{ \chi^{2} } 
+ \frac{q^{2}}{\chi^{4}}
- 8\pi P = 0,\label{Eqr}\\
&&\nonumber\\
&& \!\mathscr{C}_{\theta}{}^{\theta}\!=\!\mathscr{C}_{\varphi}{}^{\varphi}\!=\!0\hspace{0.4cm}\!\Rightarrow\!\hspace{0.4cm}
\frac{ 2\left(\chi''A + \chi'A' \right) + \chi A'' }{ 2\chi }  
- \frac{q^{2}}{\chi^{4}}
- 8\pi \tilde{P} = 0. \label{Eqte}
\end{eqnarray}
The conservation equations \eqref{ES_NLED_Eqs3} reduce
\begin{equation}\label{phi2}
P' + \frac{A'}{2A}(\rho + P) + \frac{2\chi'}{\chi} (P - \tilde{P}) = 0.
\end{equation}
Therefore, Eqs.~\eqref{Eqt}--\eqref{phi2} constitute the GR-Maxwell-FL field equations for SSS dyonic spacetimes. 

\subsection{Dyonic Simpson-Visser black-bounce in GR-Maxwell-FL gravity }

The following spacetime geometry
\begin{equation}\label{SV_BB_MaxFl}
\boldsymbol{ds^{2}} = - \left( 1 - \frac{2m}{ \sqrt{ r^{2} + q^{2} } } \right)\boldsymbol{dt^{2}} +  \left( 1 - \frac{2m}{ \sqrt{r^{2} + q^{2} } } \right)^{-1}\boldsymbol{dr^{2}}
+ \left( r^{2} + q^{2} \right)\boldsymbol{d\Omega^{2}} \quad\textup{ with }\quad q^{2} = q^{2}_{_{\mathcal{E}}} + q^{2}_{_{\mathcal{B}}},
\end{equation}
supplied with mass $m$, electric charge $q_{_{\mathcal{E}}}$ and magnetic charge $q_{_{\mathcal{B}}}$, corresponding to the electromagnetic Simpson-Visser black-bounce \eqref{B_bounce} with $a^{2}=q^{2}_{_{\mathcal{E}}} + q^{2}_{_{\mathcal{B}}}$, 
is an exact solution of GR-Maxwell-FL field equations \eqref{Eqt}$-$\eqref{phi2}
with an anisotropic fluid energy momentun tensor of the form \eqref{T_DF}
having $P=0$ and $\rho\neq0\neq\tilde{P}$ given by:
\begin{equation}\label{FL_SV}
\rho = \frac{\left(2m - \sqrt{r^{2} + q^{2}  } \right) q^{2}}{4\pi (r^{2} + q^{2})^{^{\frac{5}{2}}} }
\quad\quad\textup{and}\quad\quad
\tilde{P} = - \frac{m\!~q^{2}}{8\pi (r^{2} + q^{2})^{^{\frac{5}{2}}} },
\end{equation}
which in R-regions of the electromagnetic SV-BB spacetime ({\it i.e.}, for all $r$ such that 
$\sqrt{r^{2} + q^{2} } > 2m$)
represents an anisotropic fluid with vanishing radial pressure $P=0$,  
tangential pressure $\tilde{P}$ and  
comoving energy density $\rho$, which correspond to an  
Einstein cluster. Whereas, in T-regions 
of the electromagnetic SV-BB spacetime 
({\it i.e.}, for all $r$ such that 
$\sqrt{r^{2} + q^{2} } < 2m$) represents an 
anisotropic fluid with vanishing comoving energy density $-P=0$, 
radial pressure given by $-\rho$, and tangential pressure $\tilde{P}$.  
Thus, within the framework of general relativity, in contrast to the claim in \cite{Bronnikov2022} that SV-BB does not occur with a Maxwell electrodynamics source, the metric \eqref{SV_BB_MaxFl} describes a SV-BB spacetime geometry supported by Maxwell electrodynamics and an  
Einstein cluster in the R-regions.  
Furthermore, our solution \eqref{SV_BB_MaxFl} carries both magnetic and electric charges, whereas Refs.~\cite{Bronnikov2022,Canate2022} considered only purely magnetic black-bounces. \\
{\bf Behavior of the anisotropic fluid source of the SV-BB:} 
the electromagnetic SV-BB solution \eqref{SV_BB_MaxFl}$-$\eqref{FL_SV}, for the SV regular black hole configuration $2m \geqslant |q|$, in R-regions the SSS anisotropic fluid source of the SV regular black hole spacetime has vanishing radial pressure $P=0$, and hence corresponds to an EC with comoving energy density $\rho$ and tangential radial pressure $\tilde{P}$, such that 
\begin{equation}
\rho + P = \rho = \frac{\left(2m - \sqrt{r^{2} + q^{2} } \right)q^{2}}{4\pi (r^{2} + q^{2})^{^{\frac{5}{2}}} }  \quad\quad\textup{and}\quad\quad \rho + \tilde{P} = \frac{\left(3m - 2\sqrt{r^{2} + q^{2} } \right)q^{2}}{8\pi (r^{2} + q^{2})^{^{\frac{5}{2}}} }.
\end{equation}
This implies that the EC source of the \eqref{SV_BB_MaxFl} corresponds to exotic EC because it violates the  NEC and WEC everywhere outside the SV regular black hole, given that it has a negative comoving energy density $\rho<0$ for all $r$ in R-regions. \\ 
While in T-regions, the corresponding anisotropic fluid source of \eqref{SV_BB_MaxFl} has vanishing comoving energy density $-P=0$, and  corresponding to an exotic fluid having negative
radial pressure $-\rho$ and negative tangential pressure $\tilde{P}$ for all $r$ in T-regions, implying the violation of the NEC and WEC everywhere inside of the SV regular black hole. \\
On the other hand, for the case of the SV traversable wormhole configuration $2m < |q|$ the spacetime geommetry \eqref{SV_BB_MaxFl} only admits R-regions, and  
we obtain $ \rho + P = \rho \leqslant 0$ and $\rho + \tilde{P} \leqslant 0$ for all $r$.
Therefore, we conclude that the EC energy-momentum tensor
does not satisfy the null and weak energy conditions throughout the SV traversable wormhole spacetime.\\
{\bf Behavior of the linear electromagnetic source of the SV-BB:} For the electromagnetic SV-BB solution \eqref{SV_BB_MaxFl}$-$\eqref{FL_SV}, in both R and T-regions,
the non null components of $(T_{\alpha}{}^{\beta})\!_{_{_{\mathrm{ED}}}}$ are such that 
\begin{eqnarray}
&& (T_{t}{}^{t})\!_{_{_{\mathrm{ED}}}} = (T_{r}{}^{r})\!_{_{_{\mathrm{ED}}}}
= -\rho\!_{_{_{\mathrm{ED}}}} = P\!_{_{_{\mathrm{ED}}}}
= - \frac{q^{2}}{8\pi(r^{2} + q^{2})^{^{2}}}, \label{EDttSV}\\
&& (T_{\theta}{}^{\theta})\!_{_{_{\mathrm{ED}}}} =  (T_{\varphi}{}^{\varphi})\!_{_{_{\mathrm{ED}}}}
= \tilde{P}\!_{_{_{\mathrm{ED}}}}
= \frac{q^{2}}{8\pi(r^{2} + q^{2})^{^{2}}}.\label{EDthephSV}
\end{eqnarray}  %
Then, we can note that 
\begin{equation}
\rho\!_{_{_{\mathrm{ED}}}} > 0, \quad\quad \rho\!_{_{_{\mathrm{ED}}}} + P\!_{_{_{\mathrm{ED}}}} = 0 \quad\quad\textup{and}\quad\quad \rho\!_{_{_{\mathrm{ED}}}} + \tilde{P}\!_{_{_{\mathrm{ED}}}} > 0   \quad\quad \textup{ for all }\quad\quad  r\in\mathbb{R},
\end{equation}
indicating that the electromagnetic source of the SV-BB spacetime satisfies the null and weak energy conditions everywhere in the SV-BB spacetime. \\
{\bf Avoiding the Penrose singularity theorem:} If we consider the total source of the electromagnetic SV regular black hole spacetime, {\it i.e.,} $(T_{\alpha}{}^{\beta})\!_{_{_{e\!f\!f\!}}}=(T_{\alpha}{}^{\beta})\!_{_{_{\mathrm{ED}}}} + (T_{\alpha}{}^{\beta})\!_{_{_{\mathrm{FL}}}}$, and using the notation $\rho\!_{_{_{e\!f\!f}}}$, $P\!_{_{_{e\!f\!f}}}$ and $\tilde{P}\!_{_{_{e\!f\!f}}}$ to represented the  comoving energy density,  radial pressure and  tangential pressure of the total source of the regular black hole, respectively, within the SV regular black hole configuration $2m \geqslant |q|$, for the R-regions we obtain
\begin{eqnarray}
&&\rho\!_{_{_{e\!f\!f}}} = - (T_{t}{}^{t})\!_{_{_{\mathrm{ED}}}}
+ \rho = \frac{\left(4m - \sqrt{r^{2} + q^{2} } \right)q^{2}}{8\pi (r^{2} + q^{2})^{^{\frac{5}{2}}} }, \label{eff1}\\
&&\rho\!_{_{_{e\!f\!f}}} + P\!_{_{_{e\!f\!f}}} = -(T_{t}{}^{t})\!_{_{_{\mathrm{ED}}}} + (T_{r}{}^{r})\!_{_{_{\mathrm{ED}}}} + \rho + P = \rho = \frac{\left(2m - \sqrt{r^{2} + q^{2}}\right)q^{2}}{4\pi (r^{2} + q^{2})^{^{\frac{5}{2}}} },  \label{eff2}\\
&&\rho\!_{_{_{e\!f\!f}}} + \tilde{P}\!_{_{_{e\!f\!f}}}  = -(T_{t}{}^{t})\!_{_{_{\mathrm{ED}}}} + (T_{\theta}{}^{\theta})\!_{_{_{\mathrm{ED}}}} + \rho + \tilde{P}
= \frac{3m\!~q^{2}}{8\pi (r^{2} + q^{2})^{^{\frac{5}{2}}} }. \label{eff3}
\end{eqnarray}
Therefore, it follows that the NEC and WEC are not satisfied in the region outside of the SV regular black hole, since $\rho\!_{_{_{e\!f\!f}}} + P\!_{_{_{e\!f\!f}}} < 0$ for all $r$ in R-regions. 
While, in T-Regions 
\begin{eqnarray}
&&\rho\!_{_{_{e\!f\!f}}} = - P - (T_{r}{}^{r})\!_{_{_{\mathrm{ED}}}} = - (T_{r}{}^{r})\!_{_{_{\mathrm{ED}}}}
= \frac{q^{2}}{8\pi (r^{2} + q^{2})^{^{2}}}
\label{eff1T}\\
&&\rho\!_{_{_{e\!f\!f}}} + P\!_{_{_{e\!f\!f}}} = -(T_{r}{}^{r})\!_{_{_{\mathrm{ED}}}} + (T_{t}{}^{t})\!_{_{_{\mathrm{ED}}}} - P - \rho = - \rho = - \frac{\left(2m - \sqrt{r^{2} + q^{2}} \right)q^{2}}{4\pi (r^{2} + q^{2})^{^{\frac{5}{2}}} },  \label{eff2T} 
\end{eqnarray}
showing that for the SV regular black hole configuration, the NEC and WEC are violated  
everywhere inside the event horizon since $\rho\!_{_{_{e\!f\!f}}} + P\!_{_{_{e\!f\!f}}} < 0$ for all $r$ in T-regions.
In conclusion, the total source becomes exotic (violating the WEC inside and outside of the event horizon of the SV regular black hole spacetime).
Hence, the local energy density of the anisotropic fluid 
plus ED, $\rho^{^{(loc)}}_{\!_{_{eff}}} = (T_{\mu\nu})\!_{_{_{e\!f\!f}}} k^{\mu}k^{\nu}$,
for every timelike vector field $\boldsymbol{k}= k^{\mu}\partial_{\mu}$,
is not positive definite in the whole SV regular black hole spacetime.
Thus, the Simpson-Visser regular black hole as an exact solution of GR-Maxwell-FL
does not satisfy the WEC, with the EC being responsible for the violation of WEC 
in the R-regions of the SV regular black hole, and therefore the gravitational collapse leading to the Simpson-Visser regular black hole evades the singularity theorem of Penrose \cite{Penrose}. See Appendix for details.  

\subsection{Dyonic Ellis-Bronnikov traversable wormhole}

For $m=0$, the line element \eqref{SV_BB_MaxFl} reduces
to the following static and spherically symmetric line element,
\begin{equation}\label{EllisWH}
\boldsymbol{ds^{2}} = - \boldsymbol{dt^{2}} +  \boldsymbol{dr^{2}}
+ \left( r^{2} + q^{2} \right)\boldsymbol{d\Omega^{2}} \quad\textup{ with }\quad q^{2} = q^{2}_{_{\mathcal{E}}} + q^{2}_{_{\mathcal{B}}}.
\end{equation}
This line element is still an exact solution of GR-Maxwell-FL with
fluid determined by
\begin{equation}
\rho = - \frac{q^{2}}{4\pi (r^{2} + q^{2})^{^{2}}},
\quad\quad
P = 0\quad\quad\textup{and}\quad\quad \tilde{P} = 0,
\end{equation}
representing a dust fluid ({\it i.e.,} a pressureless perfect fluid). This solution corresponds to the dyonic Ellis-Bronnikov traversable wormhole, which has recently been discussed and generalized in Ref. \cite{Canate2024}.

\section{Uniqueness of the Simpson-Visser black-bounce solution for the Einstein-Maxwell theory coupled to an exotic Einstein cluster} \label{Uniqueness}
According to Refs. \cite{Bronnikov2022,Canate2022,Rodrigues2023} the metric ansatz that describes a static and spherically symmetric black-bounce spacetime of the Simpson-Visser type is written as,
\begin{equation}\label{SSSmet_ansatz}
\boldsymbol{ds^{2}} =  -  A(r)\boldsymbol{dt^{2}} + \frac{\boldsymbol{dr^{2}} }{A(r)}  + \chi^{2}(r) \boldsymbol{d\Omega^{2}}, \quad\quad\textup{ with }\quad\quad \chi(r) = \sqrt{r^{2} + a^{2}}.
\end{equation}
We will consider the GR-Maxwell theory coupled to an Einstein cluster 
described by an energy-momentum tensor of the form \eqref{T_DF} with $P=0$, {\it i.e.,} an anisotropic fluid with vanished radial pressure $(T_{\alpha}{}^{\beta})\!_{_{_{\mathrm{FL}}}} = \textup{diag}(-\rho, 0, \tilde{P}, \tilde{P})$.\\
Assuming the validity of Einstein's field equations \eqref{Eqt}$-$\eqref{phi2},
the difference of Eqs. \eqref{Eqt} and \eqref{Eqr} leads to 
\begin{equation}\label{rho_sol}
\rho = - \frac{1}{4\pi}\frac{\chi''}{\chi}A.
\end{equation}
Substituting $P=0$ and $\rho$ from Eq. \eqref{rho_sol} into \eqref{phi2} and solving for $\tilde{P}$ we obtain
\begin{equation}\label{P_sol}
\tilde{P} = - \frac{1}{16\pi}\frac{\chi''}{\chi'}A'.
\end{equation}
Using $\chi(r) = \sqrt{r^{2} + a^{2}}$ and replacing $\rho$ from \eqref{rho_sol}, $\tilde{P}$ from \eqref{P_sol}, into \eqref{Eqte} and solving for $A(r)$, we get 
\begin{equation}\label{A_sol}
A(r) = \frac{ q^{2} - a^{3}c_{1} }{a^{2}}  
 +  \frac{a^{2} c_{1} }{2\sqrt{r^{2} + a^{2}}} \ln\left[ \frac{(\sqrt{r^{2} + a^{2}} + a )^{2}}{r^{2}} \right] 
 + \frac{c_{2}}{ \sqrt{r^{2} + a^{2}} }.
\end{equation}
Now, if we considering  $\mathscr{C}_{\alpha}{}^{\beta} \equiv G_{\alpha}{}^{\beta} - 8\pi [ (T_{\alpha}{}^{\beta})\!_{_{_{\mathrm{ED}}}} + (T_{\alpha}{}^{\beta})\!_{_{_{\mathrm{FL}}}}]$ and evaluating $\mathscr{C}_{t}{}^{t}$ and $\mathscr{C}_{r}{}^{r}$, yields
\begin{equation}
\mathscr{C}_{t}{}^{t} = \mathscr{C}_{r}{}^{r} =  \frac{ q^{2} - a^{2} -c_{1}a^{3} }{ a^{2}\left(r^{2} + a^{2} \right) },   
\end{equation}
which implies that the field equations  $\mathscr{C}_{t}{}^{t}=0=\mathscr{C}_{r}{}^{r}$ are satisfied as long as 
\begin{equation}
c_{1} = \frac{ q^{2} - a^{2}}{ a^{3} }. 
\end{equation}
Thus, Eq. \eqref{A_sol} reduces to
\begin{equation}\label{A_solred}
A(r) =   1  + \frac{c_{2}}{ \sqrt{r^{2} + a^{2}} } +  \frac{q^{2} - a^{2} }{2a\sqrt{r^{2} + a^{2}}} \ln\!\!\left[ \frac{(\sqrt{r^{2} + a^{2}} + a)^{2}}{r^{2}} \right],
\end{equation}
whereas $c_{2}$ is fixed as $c_{2}=-2m$, where $m$ is a parameter that stands for the ADM mass of the spacetime solution, to ensure that the solution recovers the Schwarzschild limit. Therefore we arrive to the following $A(r)$ function  
\begin{equation}\label{A_solred}
A(r) =   1 -  \frac{2m}{ \sqrt{r^{2} + a^{2}} } +  \frac{q^{2} - a^{2} }{2a\sqrt{r^{2} + a^{2}}} \ln\!\!\left[ \frac{(\sqrt{r^{2} + a^{2}} + a)^{2}}{r^{2}} \right],
\end{equation}
the metric \eqref{SSSmet_ansatz} becomes
\begin{eqnarray}\label{SSSmet_gensol}
\boldsymbol{ds^{2}} &=&  -  \left( 1 - \frac{2m}{ \sqrt{r^{2} + a^{2} } } +  \frac{q^{2} - a^{2} }{2a\sqrt{r^{2} + a^{2}}} \ln\left[ \frac{(\sqrt{r^{2} + a^{2}} + a)^{2}}{r^{2}} \right] \right)\boldsymbol{dt^{2}}  \nonumber \\
 && + \left( 1 - \frac{2m}{ \sqrt{r^{2} + a^{2} } } +  \frac{q^{2} - a^{2} }{2a\sqrt{r^{2} + a^{2}}} \ln\left[ \frac{(\sqrt{r^{2} + a^{2}} + a)^{2}}{r^{2}} \right] \right)^{\!-1}\boldsymbol{dr^{2}} + (r^{2} + a^{2}) \boldsymbol{d\Omega^{2}}.
\end{eqnarray}
While \eqref{rho_sol} and \eqref{P_sol} reduce, respectively, to
\begin{eqnarray}
\rho &=& -\frac{a^{2}}{4\pi (r^{2} + a^{2})^{2}} \left( 1 - \frac{2m}{ \sqrt{ r^{2} + a^{2} } } +  \frac{q^{2} - a^{2} }{2a\sqrt{r^{2} + a^{2}}} \ln\left[ \frac{( \sqrt{r^{2} + a^{2}} + a)^{2}}{r^{2}} \right]
  \right), \label{rho_solg}\\
\tilde{P} &=& \frac{a^{2}}{16\pi (r^{2} + a^{2})^{2}}  \left\{ - \frac{2m}{\sqrt{r^{2} + a^{2}}}  + \frac{q^{2} - a^{2}}{r^{2}} \left( 1 + \frac{ r^{2} }{2a \sqrt{r^{2} + a^{2}} }\ln\left[\frac{(\sqrt{r^{2} + a^{2}} + a)^{2}}{r^{2}} \right]   \right)\right\}. \label{P_solg}
\end{eqnarray}
Since, $\lim\limits_{a\rightarrow0} \frac{1}{2a}\ln\!\!\left[ \frac{(\sqrt{r^{2} + a^{2}} + a)^{2}}{r^{2}} \right]=\frac{1}{r}$, it follows that as the parameter $a$ approaches zero the quantities \eqref{rho_solg} and \eqref{P_solg} disappear ($\rho =  0 = \tilde{P}$), whereas the line element \eqref{SSSmet_gensol} simplifies to the Reissner-Nordstr\"{o}m black hole metric.  
On the other hand, although equation \eqref{SSSmet_gensol} is the most general spacetime solution for the SSS black-bounce ansatz \eqref{SSSmet_ansatz} of the GR-Maxwell theory coupled to an Einstein cluster, this solution fails to represent a black-bounce spacetime due to the presence of a curvature singularity at $r=0$.
However, spacetime geometry \eqref{SSSmet_gensol} represents a novel SSS-AF black hole solution, with a curvature singularity at the center of symmetry ($r=0$), which generalize to the Reissner-Nordstr\"{o}m black hole solution. 
In order to produce a regular black hole solution for all $r$ values it is necessary that $a^{2} = q^{2}$, and this simplifies \eqref{SSSmet_gensol} to the dyonic Simpson-Visser black bounce metric
\begin{equation}\label{SSSmet_gen}
\boldsymbol{ds^{2}} = - \left( 1 - \frac{2m}{ \sqrt{r^{2} + q^{2}} } \right)\boldsymbol{dt^{2}} + \left( 1 - \frac{2m}{ \sqrt{r^{2} + q^{2}} } \right)^{\!-1}\boldsymbol{dr^{2}} + (r^{2} + a^{2}) \boldsymbol{d\Omega^{2}},
\end{equation}
whereas \eqref{rho_solg} and \eqref{P_solg} take the form presented in \eqref{FL_SV}, but now the resulting  solution can not be reduced to the Reissner-Nordstr\"{o}m black hole metric.  
This offers a simple explanation for why the Simpson-Visser black-bounce, which can be introduced as an electromagnetically charged solution of the general relativity theory, does not reduce to the Reissner-Nordstr\"{o}m metric. 
In contrast, the purely magnetic Simpson-Visser black-bounce solution \eqref{Qm_SV_BB} within the GR-NLED-SF framework \cite{Bronnikov2022,Canate2022,Rodrigues2023}, cannot reduce to the Reissner-Nordstr\"{o}m metric because the non linear electrodynamics model $\mathcal{L}(\mathcal{F})$ given by Eq. \eqref{U_L_SV}, which is the source of the purely magnetic Simpson-Visser black-bounce, is not physical since it does not reduce to standard Maxwell electrodynamics in the weak-field limit. 

\section{Conclusion}

In this work, we have presented, for the first time, an electromagnetic Simpson-Visser black-bounce spacetime within the framework of general relativity coupled with Maxwell electrodynamics and an exotic Einstein cluster. 
In contrast to the black-bounce spacetimes constructed in \cite{Bronnikov2022,Canate2022,Rodrigues2023}, 
which can support only magnetic charge, our SV-BB solution has both magnetic and electric charges.
Furthermore, our construction of the dyonic SV-BB is within the framework of Maxwell electrodynamics, unlike \cite{Bronnikov2022,Canate2022,Rodrigues2023,Junior2025} where nonlinear electrodynamics models (which are physically irrelevant, since in the weak-field limit they do not reduce to standard classical electromagnetism) are used to construct dyonic black-bounces.
The traversable-wormhole configuration, however, requires an exotic anisotropic fluid of the Einstein cluster type as a source in order to support the traversable wormhole structure.
The presence of this exotic EC implies that the weak energy condition is violated in the electromagnetic Simpson-Visser black bounce spacetime, and consequently allows the Penrose singularity theorem to be evaded.

\section*{Appendix}

In this appendix, we provide the explicit form of the GR-NLED-SF field equations satisfied by the metric ansatz \eqref{SSSmet} and purely magnetic configuration $\mathcal{F} = \frac{q^{2}_{_{\mathcal{B}}}}{2\chi^{4}}$. Thus, for this case, the non-null components of the Einstein tensor are \eqref{GabSSS}. 
The nonvanishing components of the energy-momentum tensor of self-interacting scalar field are,
\begin{eqnarray}
&&8\pi (T_{t}{}^{t})\!_{_{_{S \! F }}} = 8\pi (T_{\theta}{}^{\theta})\!_{_{_{S \! F }}} = 8\pi (T_{\varphi}{}^{\varphi})\!_{_{_{S \! F }}} = - \frac{1}{4}A\phi'^{2} - \mathscr{U}, \label{Ett} \\
&&8\pi (T_{r}{}^{r})\!_{_{_{S \! F }}} = \frac{1}{4}A\phi'^{2} - \mathscr{U}. \label{Err}
\end{eqnarray}
The  energy-momentum tensor components  for NLED, for the SSS spacetime with metric \eqref{SSSmet}, the purely magnetic field  \eqref{Fab_SSS_pmf}, and a generic Lagrangian density $\mathcal{L}(F)$, are given by
\begin{eqnarray}
8\pi (T_{t}{}^{t})\!_{_{_{N \! L \! E \! D}}} = 8\pi (T_{r}{}^{r})\!_{_{_{N \! L \! E \! D}}} =  -2\mathcal{L}, 
\quad 8\pi (T_{\theta}{}^{\theta})\!_{_{_{N \! L \! E \! D}}} = 8\pi (T_{\varphi}{}^{\varphi})\!_{_{_{N \! L \! E \! D}}} =  2(2\mathcal{F}\mathcal{L}_{\mathcal{F}} - \mathcal{L}) \quad \textup{ with } \quad \mathcal{F} = \frac{q^{2}_{_{\mathcal{B}}}}{2\chi^{4}}.
\end{eqnarray}
Substituting the above components in the field equations written as
$\mathscr{C}_{\alpha}{}^{\beta} = G_{\alpha}{}^{\beta} - 8\pi [ (T_{\alpha}{}^{\beta})\!_{_{_{S \! F }}} + (T_{\alpha}{}^{\beta})\!_{_{_{N \! L \! E \! D}}} ] = 0$, 
we obtain the GR-NLED-SF field equations for the metric ansatz \eqref{SSSmet} and the purely magnetic field \eqref{Fab_SSS_pmf}, take the form:
\begin{eqnarray}
&&\!\mathscr{C}_{t}{}^{t}\!=\!0\hspace{0.4cm}\!\Rightarrow\!\hspace{0.4cm}
\frac{ \left( 2\chi\chi'' + \chi'^{2} \right)A + \chi\chi' A' - 1  }{ \chi^{2} } + \frac{1}{4}A\phi'^{2} + \mathscr{U} + 2\mathcal{L} =0, \label{Eqt_NDSF}
\\
&&\nonumber\\
&&\!\mathscr{C}_{r}{}^{r}\!=\!0\hspace{0.4cm}\!\Rightarrow\!\hspace{0.4cm} \frac{ \chi'^{2}A + \chi\chi' A' - 1  }{ \chi^{2} } - \frac{1}{ 4 }A\phi'^{2} + \mathscr{U} + 2\mathcal{L} = 0, \label{Eqr_NDSF}
\\
&&\nonumber\\
&&\!\mathscr{C}_{\theta}{}^{\theta}\!=\!0\hspace{0.4cm}\!\Rightarrow\!\hspace{0.4cm} \frac{ 2\left(\chi''A + \chi'A' \right) + \chi A'' }{ 2\chi } + \frac{1}{4}A\phi'^{2} + \mathscr{U} - 2(2\mathcal{F}\mathcal{L}_{\mathcal{F}} - \mathcal{L})  =0. \label{EqteGR_NDSF}
\end{eqnarray}
The equation of motion \eqref{FEq_sf} takes the form,
\begin{equation}\label{phi2_NDSF}
\left( \chi^{2}A\phi'\right)' - 2 \chi^{2}\dot{\mathscr{U}}  = 0.
\end{equation}


{\bf Avoiding the black-hole singularity problem in general relativity:}
The Penrose singularity theorem \cite{Penrose}, and its modern variants and extensions (see for instance \cite{Senovilla}), demonstrates that in a gravitational collapse with the assumption that the Einstein field equations hold,
once a closed trapped surface $\mathcal{S}$ is formed, 
which describes the inner region of an black hole event horizon, 
then, in some region of spacetime contained in the causal future $\mathcal{J}^{+}(\mathcal{S})$ of $\mathcal{S}$, at least
one of the following must hold in order to avoid a spacetime singularity,
\begin{itemize}
\item The weak energy condition (WEC) is violated.
\item Global hyperbolicity breaks down.
\end{itemize}

{\bf Null and weak energy conditions in GR:}
For an energy-momentum tensor $T_{\mu\nu}$,
the null energy condition (NEC) stipulates that for every null vector, $n^{\alpha}$,
we obtain $T_{\mu\nu}n^{\mu}n^{\nu}\geq0$.
Following \cite{WEC}, for a diagonal energy-momentum tensor $\{T_{\alpha\beta}\} = diag\left( T_{00},T_{11},T_{22},T_{33} \right)$ 
in the spacetime coordinates $(x^{0}, x^{1}, x^{2}, x^{3})$
where $x^{0}$ is the temporal coordinate and 
$(x^{1}, x^{2}, x^{3})$ the spatial coordinates, 
which can be conveniently written as,
\begin{equation}\label{diagonalEab}
T_{\alpha}{}^{\beta} = - \rho \hskip.05cm \delta_{\alpha}{}^{0}\delta_{0}{}^{\beta} + P_{_{(1)}} \hskip.05cm \delta_{\alpha}{}^{1}\delta_{1}{}^{\beta} + P_{_{(2)}} \hskip.05cm \delta_{\alpha}{}^{2}\delta_{2}{}^{\beta} + P_{_{(3)}} \hskip.05cm \delta_{\alpha}{}^{3}\delta_{\varphi}{}^{3} ,
\end{equation}
where $\rho$ can be interpreted as the rest energy density of the matter,
 whereas $P_{_{(1)}}$, $P_{_{(2)}}$ and $P_{_{(3)}}$ are respectively the pressures along the $x^{1}$, $x^{2}$ and $x^{3}$ directions. In terms of (\ref{diagonalEab}) the NEC implies:
\begin{equation}\label{NEC_Eab}
\rho + P_{_{(a)}} \geq0 \quad\quad\textup{with}\quad\quad  a = \{ 1, 2, 3 \}.
\end{equation}
The weak energy condition (WEC) states that for any timelike vector $\boldsymbol{k} = k^{\mu}\partial_{\mu}$, ({\it i.e.,} $k_{\mu}k^{\mu}<0$), the energy-momentum tensor  obeys the inequality
$T_{\mu\nu}k^{\mu}k^{\nu} \geq 0$. This
means that the function
$\rho^{^{(loc)}} = T_{\mu\nu}k^{\mu}k^{\nu}$ (local energy density),
which represents the energy density of matter as measured by an observer whose 4-velocity is $\boldsymbol{k}$, is a non-negative quantity.
For an energy-momentum tensor of the form \eqref{diagonalEab}, 
the WEC is satisfied if and only if,
\begin{equation}\label{WEC}
\rho = - T_{0}{}^{0} \geq0, \quad\quad\quad\quad \rho + P_{_{(a)}} \geq0 \quad\textup{with}\quad  a = \{ 1, 2, 3 \}.
\end{equation}

\begin{itemize}
\item {\bf WEC for a self-interacting scalar field $(T_{\alpha}{}^{\beta})\!_{_{_{S \! F }}}$}

For the generic SSS spacetime metric \eqref{SSSmet}, evaluating Eqs. \eqref{E_SF}, \eqref{Ett} and \eqref{Err}, on the R-regions $A(r)>0$, given that for this instance $\{ x^{0}=t, x^{1}=r, x^{2}=\theta, x^{3}=\varphi \}$, we obtain 
\begin{eqnarray}
&& 8\pi \rho\!_{_{_{S \! F }}} = - 8\pi\tilde{P}\!_{_{_{S \! F }}}
=  \frac{1}{4}A\phi'^{2} + \mathscr{U}, \quad\quad\textup{ with }\quad\quad  \rho\!_{_{_{S \! F }}} = - (T_{t}{}^{t})\!_{_{_{S \! F }}}, \quad\quad \tilde{P}\!_{_{_{S \! F }}} = (P_{\theta})\!_{_{_{S \! F }}} = (P_{\varphi})\!_{_{_{S \! F }}},
\\
&& 8\pi P\!_{_{_{S \! F }}} = \frac{1}{4}A\phi'^{2} - \mathscr{U}, \quad\quad\textup{with}\quad\quad P\!_{_{_{S \! F }}} = (P_{r})\!_{_{_{S \! F }}},
\end{eqnarray}
since $\rho\!_{_{_{S \! F }}} + \tilde{P}\!_{_{_{S \! F }}}=0$, in R-regions
the tensor $(T_{\alpha}{}^{\beta})\!_{_{_{S \! F }}}$ satisfies the WEC if,
\begin{equation}\label{WEC_SF}
8\pi \rho\!_{_{_{S \! F }}} = \frac{1}{4}A\phi'^{2} + \mathscr{U} \geq0, \quad 8\pi\rho\!_{_{_{S \! F }}} + 8\pi P\!_{_{_{S \! F }}}  = \frac{1}{2}A\phi'^{2}\geq0 \quad\textup{for}\quad A(r)>0.
\end{equation}
While, in T-regions of the metric ansatz \eqref{SSSmet}, given that for this instance $\{ x^{0}=r, x^{1}=t, x^{2}=\theta, x^{3}=\varphi \}$, we obtain
\begin{eqnarray}
&&  8\pi P\!_{_{_{S \! F }}} =  8\pi\tilde{P}\!_{_{_{S \! F }}}
=  - \frac{1}{4}A\phi'^{2} - \mathscr{U}, \quad\quad\textup{with}\quad\quad P\!_{_{_{S \! F }}} = (P_{t})\!_{_{_{S \! F }}}, \quad\quad \tilde{P}\!_{_{_{S \! F }}} = (P_{\theta})\!_{_{_{S \! F }}} = (P_{\varphi})\!_{_{_{S \! F }}},
\\
&& 8\pi \rho\!_{_{_{S \! F }}} = - \frac{1}{4}A\phi'^{2} + \mathscr{U} \quad\quad\textup{with}\quad\quad \rho\!_{_{_{S \! F }}} = - (T_{r}{}^{r})\!_{_{_{S \! F }}}, \quad\quad. 
\end{eqnarray}
Therefore in T-regions, the tensor $(T_{\alpha}{}^{\beta})\!_{_{_{S \! F }}}$ satisfies the WEC if,
\begin{equation}\label{WEC_SF_Tr}
8\pi \rho\!_{_{_{S \! F }}} = -\frac{1}{4}A\phi'^{2} + \mathscr{U} \geq0, \quad 8\pi\rho\!_{_{_{S \! F }}} + 8\pi P\!_{_{_{S \! F }}}  = -\frac{1}{2}A\phi'^{2}\geq0 \quad\textup{for}\quad A(r)<0.
\end{equation}
\item {\bf WEC for the nonlinear electromagnetic field $(T_{\alpha}{}^{\beta})\!_{_{_{N \! L \! E \! D}}}$ for static and spherically symmetric purely magnetic configuration $\mathcal{F} = \frac{q^{2}_{_{\mathcal{B}}}}{2\chi^{4}}$ }

For the generic SSS spacetime metric \eqref{SSSmet} the non null components of $(T_{\alpha}{}^{\beta})\!_{_{_{N \! L \! E \! D}}}$ for the purely magnetic field \eqref{Fab_SSS_pmf}, are
\begin{eqnarray}
&& 8\pi(T_{t}{}^{t})\!_{_{_{N \! L \! E \! D}}} = 8\pi(T_{r}{}^{r})\!_{_{_{N \! L \! E \! D}}} = - 2\mathcal{L},  
\label{EDttNL}\\
&& 8\pi(T_{\theta}{}^{\theta})\!_{_{_{N \! L \! E \! D}}} = 8\pi(T_{\varphi}{}^{\varphi})\!_{_{_{N \! L \! E \! D}}} = 2(2\mathcal{F}\mathcal{L}_{\mathcal{F}} - \mathcal{L}), \label{EDthephNL}
\end{eqnarray} 
Hence, in both R-regions and T-regions, yields 
\begin{eqnarray}%
&&8\pi \rho\!_{_{_{N \! L \! E \! D}}} = - 8\pi P\!_{_{_{N \! L \! E \! D}}} =  2\mathcal{L}, \quad\quad\textup{ with }\quad\quad P\!_{_{_{N \! L \! E \! D}}} = (P_{r})\!_{_{_{N \! L \! E \! D}}},
\nonumber \\
&& 8\pi\tilde{P}\!_{_{_{N \! L \! E \! D}}} =  2(2\mathcal{F}\mathcal{L}_{\mathcal{F}} - \mathcal{L}), \quad\quad\textup{ with }\quad\quad \tilde{P}\!_{_{_{N \! L \! E \! D}}} = (P_{\theta})\!_{_{_{N \! L \! E \! D}}} = (P_{\varphi})\!_{_{_{N \! L \! E \! D}}},
\end{eqnarray}
since $\rho\!_{_{_{N \! L \! E \! D}}} + P\!_{_{_{N \! L \! E \! D}}} = 0$, the tensor $(T_{\alpha}{}^{\beta})\!_{_{_{N \! L \! E \! D}}}$ satisfies the WEC if,
\begin{equation}\label{WEC_NLED}
8\pi \rho\!_{_{_{N \! L \! E \! D}}} =  2\mathcal{L}\geq0, \quad\quad  8\pi\rho\!_{_{_{N \! L \! E \! D}}} + 8\pi \tilde{P}\!_{_{_{N \! L \! E \! D}}} =  4\mathcal{F}\mathcal{L}_{\mathcal{F}}\geq0.
\end{equation}
\item {\bf  WEC for the effective energy-momentum tensor $(T_{\alpha}{}^{\beta})\!_{_{_{e\!f\!f\!}}}=(T_{\alpha}{}^{\beta})\!_{_{_{S \! F }}} + (T_{\alpha}{}^{\beta})\!_{_{_{N \! L \! E \! D}}}$ for static and spherically symmetric purely magnetic configuration $\mathcal{F} = \frac{q^{2}_{_{\mathcal{B}}}}{2\chi^{4}}$ } 

In R-regions of the generic SSS spacetime metric \eqref{SSSmet} 
\begin{eqnarray}
&&8\pi \rho\!_{_{_{e\!f\!f\!}}} = \frac{1}{4} A \phi'^{2} + \mathscr{U} + 2\mathcal{L}, \\
&&8\pi P\!_{_{_{e\!f\!f\!}}} = 8\pi(P_{r})\!_{_{_{e\!f\!f\!}}} = \frac{1}{ 4 } A \phi'^{2} - \mathscr{U} - 2\mathcal{L}, \\
&&8\pi \tilde{P}\!_{_{_{e\!f\!f\!}}} = 8\pi(P_{\theta})\!_{_{_{e\!f\!f\!}}} = 8\pi (P_{\varphi})\!_{_{_{e\!f\!f\!}}} = -\frac{1}{4} A \phi'^{2} - \mathscr{U} + 2(2\mathcal{F}\mathcal{L}_{\mathcal{F}} - \mathcal{L}),
\end{eqnarray}
So, for $A(r)>0$, the tensor $(T_{\alpha}{}^{\beta})\!_{_{_{e\!f\!f\!}}}$ satisfies the WEC if,
\begin{eqnarray}
&&8\pi \rho\!_{_{_{e\!f\!f\!}}} = \frac{1}{4}A\phi'^{2} + \mathscr{U} + 2\mathcal{L} \geq 0, \\
\nonumber\\
&& 8\pi \rho\!_{_{_{e\!f\!f\!}}} + 8\pi P\!_{_{_{e\!f\!f\!}}} = \frac{1}{2} A \phi'^{2} \geq 0, \label{WEC_effect}\\
\nonumber\\
&& 8\pi\rho\!_{_{_{e\!f\!f\!}}} + 8\pi \tilde{P}\!_{_{_{e\!f\!f\!}}}
= 4\mathcal{F}\mathcal{L}_{\mathcal{F}}\geq0.
\end{eqnarray}
In T-regions of the generic SSS spacetime metric \eqref{SSSmet} 
\begin{eqnarray}
&&8\pi \rho\!_{_{_{e\!f\!f\!}}} = - \frac{1}{4} A \phi'^{2} + \mathscr{U} + 2\mathcal{L}, \\
&&8\pi P\!_{_{_{e\!f\!f\!}}} = 8\pi(P_{r})\!_{_{_{e\!f\!f\!}}} = -\frac{1}{ 4 } A \phi'^{2} - \mathscr{U} - 2\mathcal{L}, \\
&&8\pi \tilde{P}\!_{_{_{e\!f\!f\!}}} = 8\pi(P_{\theta})\!_{_{_{e\!f\!f\!}}} = 8\pi (P_{\varphi})\!_{_{_{e\!f\!f\!}}} = -\frac{1}{4} A \phi'^{2} - \mathscr{U} + 2(2\mathcal{F}\mathcal{L}_{\mathcal{F}} - \mathcal{L}).
\end{eqnarray}
So, for $A(r)<0$, the tensor $(T_{\alpha}{}^{\beta})\!_{_{_{e\!f\!f\!}}}$ satisfies the WEC if,
\begin{eqnarray}
&&8\pi \rho\!_{_{_{e\!f\!f\!}}} = -\frac{1}{4}A\phi'^{2} + \mathscr{U} + 2\mathcal{L} \geq 0, \\
\nonumber\\
&& 8\pi \rho\!_{_{_{e\!f\!f\!}}} + 8\pi P\!_{_{_{e\!f\!f\!}}} = - \frac{1}{2} A \phi'^{2} \geq 0, \label{WEC_effectTr}\\
\nonumber\\
&& 8\pi\rho\!_{_{_{e\!f\!f\!}}} + 8\pi \tilde{P}\!_{_{_{e\!f\!f\!}}}
= - \frac{1}{2} A \phi'^{2} + 4\mathcal{F}\mathcal{L}_{\mathcal{F}}\geq0.
\end{eqnarray}  
\end{itemize}

%

\end{document}